\documentclass[prl,aps,showpacs,twocolumn,unsortedaddress,superscriptaddress]{revtex4}
\usepackage{graphics,bm}
\usepackage{amssymb}
\usepackage{psfrag}
\usepackage{epsfig}
\usepackage{epsf}

\begin{document}

\title{Pinning of vortices in a Bose-Einstein condensate by an optical lattice}

\author{J.W. Reijnders}
\email{jwr@science.uva.nl}
\affiliation{Institute for Theoretical Physics,
University of Amsterdam, Valkenierstraat 65, 1018 XE
Amsterdam, The Netherlands}

\author{R.A. Duine}
\email{duine@physics.utexas.edu}
\homepage{http://www.ph.utexas.edu/~duine} \affiliation{The
University of Texas at Austin, Department of Physics, 1 University
Station C1600, Austin, TX 78712-0264} \affiliation{Utrecht
University, Institute for Theoretical Physics, Leuvenlaan 4, 3584
CE Utrecht, The Netherlands}

\date{\today}

\begin{abstract} We consider the ground state of vortices in a
Bose-Einstein condensate. We show that turning on a weak optical periodic
potential leads to a transition from the triangular Abrikosov vortex
lattice to phases where the vortices are pinned by the optical potential.
We discuss the phase diagram of the system for a two-dimensional optical
periodic potential with one vortex per optical lattice cell. We also
discuss the influence of a one-dimensional optical periodic potential
on the vortex ground state. The latter situation has no analogue in
other condensed-matter systems.
\end{abstract}


\maketitle

\def\bx{{\bf x}}
\def\bk{{\bf k}}
\def\half{\frac{1}{2}}
\def\args{(\bx,t)}

{\it Introduction} --- The effects of a periodic array of pinning centers
on vortices in superconducting materials have attracted a lot of
experimental
\cite{fiory1978,baert1995,castellanos1997,martin1997,morgan1998} and
theoretical \cite{nelson1979,radzihovsky1995,blatter1994,reichhardt1997,
kolton1999,dasgupta2002,pogosov2003,zhuravlev2003} attention. Of particular
interest is the effect of the pinning potential on the melting of a vortex
lattice. The vortex lattice is known to melt via a first-order transition
in clean materials \cite{brezin1985}, whereas the presence of pinning
centers significantly enriches the phase diagram, due to the intricate
interplay between the vortex-vortex interactions, pinning potential, and
thermal and quantum fluctuations
\cite{nelson1979,radzihovsky1995,blatter1994}. At zero temperature and for
strong pinning, the system has, depending on the number of vortices per
pinning center, i.e., the filling factor, various phases where the vortices
order in a periodic array
\cite{reichhardt1997,kolton1999,dasgupta2002,zhuravlev2003}. If the pinning
potential is weakened, the pinned vortex lattice undergoes a first-order
transition to a deformed triangular Abrikosov lattice \cite{pogosov2003}.

Recently, the experimental study of vortices in superfluids has
been complemented by the experiments with rotating atomic Bose
gases \cite{matthews1999,madison2000,raman2001,hodby2002}. Within
this field it has become possible to experimentally study the
dynamics of a single vortex line in great detail
\cite{anderson2000,rosenbusch2002}, leading to an enhancement of
the theoretical interest in the dynamics of a single vortex in a
Bose-Einstein condensed atomic gas \cite{fetter2001}.

Another interesting development in the field of atomic gases is the
possibility to trap atoms in a periodic potential using a so-called
optical lattice. Here, one uses the dipole force which the atoms
experience in an off-resonant light field. Using an optical lattice
Greiner {\it et al.} \cite{greiner2002} were able to experimentally
observe the transition from a superfluid, where the atoms
are delocalized across the lattice, to a Mott-insulating state where the
atoms are localized onsite \cite{fischer1989}.

A common feature of the experiments with ultracold atomic gases is
that these systems are very clean. Therefore, vortices in a
Bose-Einstein condensate do not experience an intrinsic pinning
potential and the observed vortex lattices are triangular
Abrikosov lattices. In this Letter, we study the ground state of
vortices in a Bose-Einstein condensate. We show that turning on an
optical periodic potential leads to a transition from the
triangular Abrikosov lattice to phases where the vortices are
pinned by the optical potential. We restrict ourselves to the case
with one vortex per unit cell of the two dimensional optical
lattice.

Interestingly, the precise knowledge of the optical lattice potential
allows for a microscopic and quantitative calculation of the phase diagram,
as opposed to superconducting materials, where the pinning potential is
known only phenomenologically. In the experiments with rotating
Bose-Einstein condensates, the vortex lattices are relatively easy to
observe, which allows for a detailed experimental study of the transitions
between the various phases of the vortex lattice as one tunes the strength
of the optical potential. Moreover, applying a one-dimensional optical
lattice leads to pinning ``valleys'' instead of pinning centers, and,
therefore, to pinned phases of the vortex lattice which have no analogue in
other condensed-matter systems.

{\it Two-dimensional optical lattice} --- Our starting point is
the hamiltonian functional for the macroscopic condensate wave
function $\Psi (\bx)$, given by
\begin{eqnarray}
H[\Psi^*,\Psi]&=& \int {d \bf x}\Psi^*({\bf x}){\bigg [}
-{\hbar^2\nabla^2 \over 2 m} + {1\over2}g|\Psi({\bf x})|^2\nonumber\\
&~&~~ + V_{\rm OL}({\bf x})-\mu {\bigg ]}\Psi({\bf x})~.
\label{eq:hamiltonian}
\end{eqnarray}
Here, $m$ denotes the mass of one atom which interacts with the other
atoms via a two-body contact interaction of strength
$g=4 \pi a_s \hbar^2/m$, with $a_s$ the $s$-wave scattering length.
The optical lattice potential is given by $V_{\rm OL}({\bf x})=s E_{\rm R}
[\sin^2(qx)+\sin^2(q y)]$ with $E_{\rm R}$ the recoil energy, $q$
the wavenumber of the lattice, and $s\geq 0$ a dimensionless number
indicating the strength of the optical lattice. Note that considering
one vortex per unit cell of the optical lattice implies that we take
the two-dimensional vortex density equal to $n_{\rm v}=q^2/\pi^2$.
The chemical potential that fixes the number of atoms in the condensate
is represented by $\mu$.

In the following we consider for simplicity a condensate with
infinite extent in the $x$-$y$-plane and tightly confined in the
$z$-direction by an harmonic trap with frequency $\omega_z$. The
latter assumption allows us to neglect the curvature of the vortex
lines, and leads effectively to a condensate thickness
$d\equiv\sqrt{\pi \hbar/(m\omega_z)}$ in the $z$-direction. In the
Thomas-Fermi limit, where we neglect the kinetic energy of the
condensate atoms with respect to their mean-field interaction
energy, the global density profile of the condensate in the
optical lattice is given by $n_{\rm TF}({\bf x})=n-[V_{\rm
OL}({\bf x})-sE_{\rm R}]/g$, with $n=[\mu-sE_{\rm R}]/g$ the
average density of the condensate. To find the potential energy of
a vortex in a Bose-Einstein condensate in an optical lattice as
function of its coordinates ${(u_x,u_y)}$, we use the variational
{\it ansatz} for the condensate wave function
\begin{equation}
\Psi ({\bf x}) = \sqrt{n_{\rm TF}({\bf x})}~\Theta {\big [}{|{\bf x}-
{\bf u}|/\xi}-1{\big ]}\exp[i \phi({\bf x},{\bf u})]~,
\label{eq:vortexansatz}
\end{equation}
with $\xi=1/\sqrt{8 \pi a_s n}$ the healing length that sets the size of
the vortex core, $\phi({\bf x},{\bf u})=\arctan[(y-u_y)/(x-u_x)]$ the phase
of the vortex, and $\Theta(z)$ the unit step function. For the above
{\it ansatz} to be valid, we have assumed that the vortex core is much
smaller then an optical lattice period, $q \xi \ll 1$, and that the
strength of the potential is sufficiently weak, $s E_{\rm R} \ll \mu$.
Note that the above variational {\it ansatz} indeed describes a vortex
along the $z$-axis at position $(u_x,u_y)$ in the $x$-$y$-plane.

\begin{figure}
\psfrag{SP}{\bf\large SP}
\psfrag{HP}{\bf\large HP}
\psfrag{AB}{\bf\large AB}
\psfrag{vor}{pinning center}
\psfrag{pin}{vortex}
\psfrag{qx}{$q \xi$}
\psfrag{se}{\large $s E_R \over \mu$}
\includegraphics[width=8cm,height=5cm]{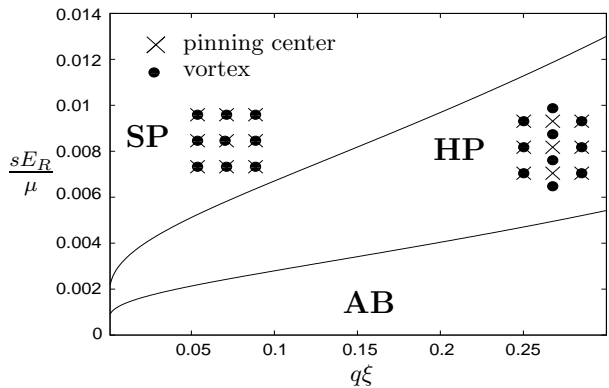}
\caption{\label{fig:phasediagram} Vortex phase diagram of a Bose-Einstein
condensate in a two-dimensional optical square lattice, with one vortex per
unit cell of the optical lattice. Three phases are relevant: a square and
fully pinned configuration (SP), a triangular configuration where half of
the vortices are located at the pinning centers (HP), and the unpinned
triangular Abrikosov vortex lattice (AB).}
\end{figure}

Substitution of the {\it ansatz} in Eq.~(\ref{eq:vortexansatz}) in
the hamiltonian functional in Eq.~(\ref{eq:hamiltonian}), and
isolating the contribution due to the presence of the vortex leads
to \cite{kevrekidis2003}
\begin{equation}
U_{\rm 2D}({\bf u})={ d \over 8 a_s}s E_{\rm R} Q(q \xi)
[\cos(2q u_x)+\cos(2q u_y)]~.
\label{eq:potential}
\end{equation}
Here, we defined $Q(z)= J_1(2 z)/(2z) +\int_1^\infty d\rho
J_0(2z\rho)/\rho$, with $J_0$ and $J_1$ the zeroth and first order
Bessel function of the first kind. It is clearly seen that the potential
energy is minimal if the vortices are located at the maxima of the
optical potential. This is expected, since at these maxima the condensate
density, and hence the kinetic energy associated with the superfluid
motion, is minimal. The expression in Eq. (\ref{eq:potential}) is the
pinning potential experienced by vortices in a condensate loaded in a
optical lattice. If the pinning potential is sufficiently strong and
we have one vortex per unit cell of the optical lattice, the ground
state is a configuration in which each vortex is trapped or pinned by
an optical lattice maximum. For a two-dimensional optical periodic
potential, this is the square pinned lattice (SP), shown in
Fig.~\ref{fig:phasediagram}.

To determine the phase diagram in detail, we calculate the total
energy per vortex for an arbitrary vortex lattice. This approach
neglects the fact that for very weak pinning potentials the
triangular Abrikosov lattice will be slightly deformed
\cite{pogosov2003}. Generally, the vortex lattice is parametrized
as follows
\begin{equation}
{\bf u}(\alpha, \beta)~=~
{\pi \over q}
\left(\begin{array}{cc}
\sqrt{1+\alpha\beta}&\alpha\\ \beta&\sqrt{1+\alpha\beta}
\end{array}\right)
\left(\begin{array}{c}
n_x\\n_y
\end{array}\right)~,~
\label{eq:unitcellparametrization}
\end{equation}
with $n_i\in\mathbb{Z}$ and $0\leq \alpha,\beta\leq {1\over 2}$. The
transformation matrix conserves the area of the unit cell, and thus
ensures that we are considering configurations with equal vortex density.
The more familiar parameters of a unit cell of a two-dimensional lattice,
the angle $\varphi$ and the ratio of the length of the sides of the
cell $\kappa=L_1/L_2$, are related to $\alpha$ and $\beta$ by
\begin{equation}
\frac{\cos \varphi}{\kappa}=\frac{(\alpha+\beta)\sqrt{1+\alpha\beta}}
{1+\alpha\beta+\alpha^2},\quad
\frac{\sin \varphi}{\kappa}=\frac{1}{1+\alpha\beta+\alpha^2}.
\label{eq:relation}
\end{equation}
The pinning energy per vortex is found by substituting Eq.
(\ref{eq:unitcellparametrization}) in Eq. (\ref{eq:potential}), summing over
all $n_i$, and dividing the result by the number of unit cells. In the
limit $n_i\rightarrow\infty$ we find
\begin{equation}
E^{\rm 2D}_{\rm pin}(\alpha,\beta)=-{ d \over 8 a_s}s E_{\rm R} Q(q \xi)
[\delta_{\beta,0}+\delta_{\alpha,0}]~.
\label{eq:pinningenergy}
\end{equation}

Our next task is to determine the interaction energy per vortex.
In two dimensions, singly-quantized vortices with equal orientation
experience a logarithmic long-range interaction
$V(r)= - 2\pi d\hbar^2n/m \log (r/\xi)$ for $r \gg \xi$ \cite{kleinertbook}.
The interaction energy $E_{\rm int}$ per vortex for an infinite
two-dimensional lattice of vortices was calculated by Campbell {\it et al.}
\cite{campbell1989}.  Cast in a dimensionless form, their result reads
\begin{eqnarray}
&& \tilde E_{\rm int} \equiv \frac{E_{\rm int}}{(\pi \hbar^2 d n/m)} =
\frac{\pi}{6} \frac{\sin
\varphi}{\kappa}
 -\log \left[ 2 \pi \left( \frac{\sin \varphi}{\kappa
 }\right)^{1\over 2}\right]\nonumber \\
&&
 -\log \big\{\Pi_{j=1}^{\infty} [ 1-
 2 e^{-2\pi j|\sin\varphi|/\kappa}\cos\left(2 \pi j \frac{\cos \varphi}{
\kappa}\right)\nonumber\\
&& +~e^{-4\pi j |\sin \varphi|/\kappa}] \big\}~.
\label{eq:interactionenergy}
\end{eqnarray}
It is important to realize that the interaction energy per vortex is
divergent for an infinite lattice, and that the above expression gives the
relative interaction energy for configurations with equal vortex density.

The expression for $E_{\rm int}$ together with the expression for the
pinning energy in Eq.~(\ref{eq:pinningenergy}) enables us to minimize
the total energy as a function of $\alpha$ and $\beta$, and to
determine the vortex lattice ground state for different strengths of the
periodical optical potential. Performing this procedure as function of the
dimensionless variables $q \xi$ and $s E_{\rm R}/\mu$ leads to three
 phases, which are physically due to the the competition between
pinning and interactions. For a very weak optical potential we find that
vortices arrange themselves on a triangular Abrikosov lattice (AB), with
$\alpha=\beta=\sqrt{1/\sqrt{3}-1/2}$, as expected. In this phase, with
${\tilde E}_{\rm int}=-1.32112$, the interactions dominate over the pinning.
On the other hand, when the pinning energy dominates over the interaction,
we find the square pinned lattice ($\alpha=\beta=0$) \cite{pogosov2003}.
This phase has ${\tilde E}_{\rm int}=-1.31053$.
  In the intermediate regime, where the pinning and interactions are
equally important, we find a phase in which half of the vortices are pinned
(HP) \cite{pogosov2003}, and the lattice has a triangular structure ($\alpha
= {1 \over 2},~\beta=0$ and ${\tilde E}_{\rm int}=-1.31800$). In the
zero-temperature phase diagram, shown in Fig.~\ref{fig:phasediagram}, the
different phases are separated by first-order phase transitions and the phase
boundaries are given by
\begin{equation}
\left( \frac{sE_{\rm R}}{\mu}\right)_{\rm AB-HP}={.00623 \over
Q(q\xi)},~
\left( \frac{sE_{\rm R}}{\mu} \right)_{\rm HP-SP}={.01494 \over Q(q\xi)}.
\label{eq:phaseboundaries}
\end{equation}

\begin{figure}
\psfrag{PT}{\bf\large PT}
\psfrag{AB}{\bf\large AB}
\psfrag{q}{$q/q_0$}
\psfrag{se}{\large $s E_R \over \mu$}
\includegraphics[width=8cm,height=5cm]{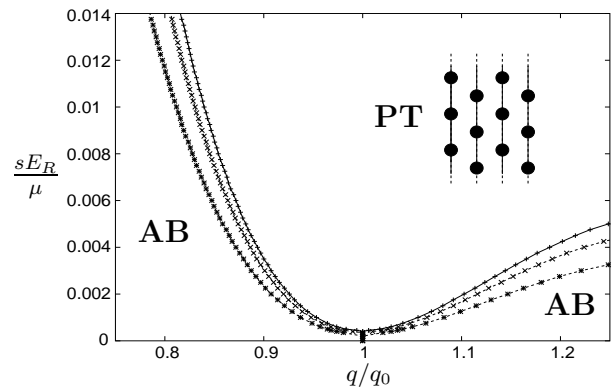}
\caption{\label{fig:1dphasediagram} Vortex phase diagram of a
Bose-Einstein condensate in a one-dimensional optical lattice. Two
phases are relevant: a pinned triangular configuration (PT), and
the unpinned triangular Abrikosov vortex lattice (AB). We
calculated the phase boundaries for $\xi=.01 \pi/q_0$ ($+$),
$\xi=.005\pi/q_0$ ($\times$) and $\xi=.001\pi/q_0$ ($\ast$). Note
that at the line $q/q_0=1$ the phases coincide. }
\end{figure}

{\it One-dimensional optical lattice} --- For a one-dimensional optical
lattice the single vortex pinning potential equals $U_{\rm 1D}({\bf u})={
d \over 8 a_s}s E_{\rm R} Q(q \xi)\cos(2q u_x)$ and hence the minima of
$U_{\rm 1D}$ act as pinning ``valleys''. The pinning energy per vortex
reads $E_{\rm pin}^{\rm 1D} = -{ d \over 8 a_s}s E_{\rm R} Q(q \xi)
\delta_{\beta,0}$ within the parametrization in
Eq.~(\ref{eq:unitcellparametrization}). Two phases are distinguished in
this case, i.e., a pinned triangular lattice (PT), shown in
Fig.~\ref{fig:1dphasediagram}, and the unpinned Abrikosov vortex lattice.
Interestingly, the interactions always favor a triangular lattice since
the vortices are allowed to arrange freely in the $y$-direction.

Consider now the case that the wavenumber of the optical lattice is such
that the AB lattice and the PT lattice coincide, i.e., $q_0 \equiv
2\pi \sqrt{3}/(3 L)$, with $L$ the intervortex distance. If this
configuration is disturbed by changing the optical lattice wavenumber
to arbitrary $q$ there will be a competition between the AB lattice
and the PT lattice. The unit cell of the latter is, for wavenumber
$q$ and at equal vortex density, described by
\begin{equation}
\frac{\cos \varphi}{\kappa}=\frac{2}{1+3 \left(
\frac{q_0}{q}\right)^4}~,\quad
\frac{\sin \varphi}{\kappa}=\frac{2\sqrt{3} \left(
\frac{q_0}{q}\right)^2}{1+3\left( \frac{q_0}{q}\right)^4}~.
\label{eq:1dparameters}
\end{equation}
The interaction energy per unit cell is found by substitution of
Eq.~(\ref{eq:1dparameters}) in Eq.~(\ref{eq:interactionenergy}).
From this we find the zero-temperature phase diagram, depicted in
Fig.~\ref{fig:1dphasediagram} for various values of the healing length.
The generic behavior is such that for given strength of the optical
lattice and for small deviations from $q_0$, the vortex lattice
stays pinned, i.e., the vortices are dragged along with the pinning
valleys.  At certain $q$ the phase boundary is crossed. Then the
vortices ``jump'' back to their original positions, forming an
Abrikosov lattice again.

{\it Discussion and conclusions} --- Since we have considered the
ground state of vortices, we have implicitly assumed that the
optical lattice is co-rotating with the Bose-Einstein condensate.
Although this is very difficult to achieve experimentally, it has,
however, recently been proposed to create effective magnetic
fields, and therefore effective rotation, by optical methods
\cite{jaksch2003,mueller2004}. We believe that the calculations
presented here are relevant for such a situation.

Of particular experimental interest are the collective modes
supported by the pinned vortex lattices. In the absence of an
optical lattice potential, the dispersion relation for the gapless
Tkachenko modes has been measured by Coddington {\it et al.}
\cite{coddington2003}. We expect that these modes acquire a gap in
the presence of a periodic optical potential due to the fact that
the translational symmetry is not broken spontaneously, but by the
optical lattice. In the case of the SP lattice this gap is easy to
calculate, since the zero-momentum Tkachenko mode corresponds in
this case to a simultaneous in-phase precession of all the
vortices around the maxima of the optical lattice. Hence the gap
is given by $\hbar \omega_{\rm p}$, with $\omega_{\rm p}$ the
precession frequency \cite{fetter2001}
\begin{equation}
\label{eq:precfreq}
  \omega_{\rm p} = \frac{\pi \hbar q^2 Q(q\xi) s E_{\rm R}}{mgn}~.
\end{equation}
In future work we intend to study the collective modes of the pinned
vortex lattices in great detail.

In this Letter we have restriced ourselves to the case of one vortex per
unit cell of the optical lattice. Another direction for future work will
be a detailed study of the pinned phases at other filling factors, where
other types of pinned vortex lattices are known to occur
\cite{reichhardt1997,kolton1999,dasgupta2002,zhuravlev2003}. Since we have
only been considering infinite lattices, we intend to study also the
effects of the finite size of the system, which may be significant
\cite{campbell1979}. With respect to this latter remark it is also
important to note that the harmonic magnetic trap used in the experiments
induces an additional feature in the density profile of the condensate
that may have important effects \cite{anglin2002}, and therefore also
requires further study.

Apart from these interesting possibilities, yet another direction
would be to consider more strongly-correlated regimes that occur
at fast rotation, and to study the effects of the periodic optical
potential on the melting of the vortex lattice \cite{sinova2002}.
One would expect that in this regime the effect of quantum
fluctuations, i.e., quantum tunneling of the vortices through the
potential barriers of the pinning potential, becomes important.
Conversely, in the limit of a strong optical periodic potential,
an interesting topic is to study the effects of the rotation on
the Mott-Insulator transition \cite{wu2002}. In conclusion,
rotating Bose gases in an optical lattice provide an interesting
system to study new quantum phases of matter
\cite{martikainen2003}, as well as phenomena known from condensed
matter in a new context.

It is a great pleasure to acknowledge Kareljan Schoutens and Henk Stoof
for useful remarks. This research was supported by the Netherlands
Organization for Scientific Research, NWO.
R.A.D. acknowledges partial support by the National Science
Foundation under grant DMR-0115947.

\end{document}